\documentclass[prd,aps,showpacs,showkeys]{revtex4-1}
\usepackage{bm,amsfonts,latexsym,amsmath,amssymb,amsbsy,graphicx}

\newcommand{\bb}{\begin{eqnarray}}
\newcommand{\ee}{\end{eqnarray}}
\newcommand{\ba}{\begin{align}}
\newcommand{\ea}{\end{align}}

\begin{document}

\title{\bf Vacuum polarization of planar Dirac fermions by a superstrong Coulomb potential}
\author{V.R. Khalilov \footnote{Corresponding author}}\email{khalilov@phys.msu.ru}
\affiliation{Faculty of Physics, Lomonosov Moscow State University, 119991,
Moscow, Russia}
\author{I.V. Mamsurov}
\affiliation{Faculty of Physics, Lomonosov Moscow State University, 119991,
Moscow, Russia}

\begin{abstract}
We study the vacuum polarization of planar charged Dirac fermions by a strong Coulomb potential.
Induced vacuum  charge density is calculated and analyzed at the subcritical and supercritical Coulomb potentials for massless and massive fermions.  For the massless case  the induced vacuum charge density is localized at the origin
when the Coulomb center charge is subcritical  while it has a power-law tail  when the Coulomb center charge is supercritical. The finite mass contribution
into the induced charge due to the vacuum polarization is small and insignificantly distorts
the Coulomb potential only at distances of order of the Compton length. The  induced vacuum charge has a  screening sign.  As is known the quantum electrodynamics vacuum becomes unstable when the Coulomb center charge is increased from subcritical to supercritical values. In the supercritical Coulomb potential the quantum electrodynamics vacuum acquires the charge due to the so-called real vacuum polarization. We calculate the real vacuum polarization charge density. Screening of the Coulomb center charge are briefly  discussed. We expect  that our results will be helpful for more deep understanding of the fundamental problem of quantum electrodynamics and can as a matter of principle  be tested in graphene with a supercritical Coulomb impurity.
\end{abstract}



\maketitle

\section{Introduction}

%PACS numbers: 73.22.Pr, 81.05.ue, 03.65.Pm

The vacuum of the quantum electrodynamics  and
the induced vacuum polarization in a strong Coulomb field produced
by a heavy atomic nucleus  have been studied a long time \cite{003,004,grrein,01,02,03,04}.
  When the nuclear charge $Z|e|$ ($e$ is the electron charge) is increased from subcritical
to supercritical values then the lowest electron energy level (in the regularized Coulomb potential) dives into the negative energy continuum and becomes a resonance with complex ``energy''$E=|E|e^{i\tau}$ signaling  the instability of the quantum electrodynamics vacuum in the supercritical range.
The  nuclear  charge $Z_{cr}|e|$ for which the lowest energy level descends to the negative-energy continuum boundary $-m$ is called the critical charge for the ground state. The critical charge is obviously related to the fine structure constant $1/137$ and  the number  $Z_{cr}\sim 170$ \cite{05}.

New interest to these problems was revived in connection with the charged impurity problem in
graphene because  charged impurities can produce the supercritical Coulomb potential due to the corresponding ``effective fine structure constant" is large.
In graphene, the electrons near the Fermi surface can be described
in terms of an effective Lorentz-invariant theory with their energy
determined by Dirac's dispersion law for massless fermions \cite{6,7,review}, which allows to consider graphene as the condensed matter analog of the quantum electrodynamics
in 2+1 dimensions \cite{datdc,ggvo}. The existence of charged Fermi quasiparticles
in graphene makes  experimentally feasible to observe the vacuum polarization in strong Coulomb field
but the massless case turn out to be rather more complicated
since an infinite number of quasi-localized resonances emerges
in the hole sector at the supercritical Coulomb potential \cite{vp11a,as11b,ggg}.

Charged impurity screening produced by massless charged fermions in graphene in terms of vacuum polarization  were investigated  in  \cite{7,review,vp11a,as11b,13a,as112,bsaso,kn11c,12,vmpvnk,yn1}.
For massless fermions the induced vacuum  charge density is localized
at the origin   in the subcritical Coulomb potential \cite{vp11a,bsaso,12}
while it  has the form  $c/r^2$, therefore, causing  a modification of the supercritical Coulomb potential \cite{vp11a,as11b}. The vacuum polarization of the massive charged fermions can also be of interest for graphene with Coulomb impurity \cite{13}. For massive fermions the vacuum polarization charge density  behaves differently from the massless ones.

%The  wonderful quantum phenomenon was revealed in \cite{15}: the induced
%current density in graphene in the field of a solenoid turns out to be a
%finite periodic function of the magnetic flux.
%The  induced polarization current in the QED$_{2+1}$ with an Aharonov--Bohm  potential
%for  massive and massless charged fermions was studied in \cite{kh1}.
%The induced electric current  due to vacuum polarization in the AB potential
%was observed in \cite{34a} in ``a quantum-tunneling system using two-dimensional ionic
%structures in a linear Paul trap''. It is worth to note that the vacuum charge density
% is induced by the homogeneous background magnetic field in
%the massive and massless ${\rm QED}_{2+1}$ \cite{khmam}.

The dynamics of charged fermions in a Coulomb potential  is governed by a singular Dirac Hamiltonian that requires the supplementary definition in order for it to be treated
as a self-adjoint quantum-mechanical operator. So, at first we need
to determine the self-adjoint Dirac Hamiltonians and then to construct the correct Green function
of the Dirac equation in a singular Coulomb potential.
The self-adjoint Dirac Hamiltonians are not unique and can be specified by
a self-adjoint extension parameter which implies
 additional nontrivial boundary conditions on the wave functions at the origin \cite{17}.

%The self-adjoint extension
%method was used  to determine bound states of massive fermions
%in the Aharonov--Bohm-like fields \cite{as0,sil,kh2}.

%For example,  in an Aharonov--Bohm field the magnetic flux within the interior of the vortex
%determines the effective Hamiltonian outside it; the extensions can be parameterized
%by  boundary conditions  at the origin
%and different choices lead to inequivalent physical cases \cite{phg}.
%We can determine the self-adjoint
%extension parameter in terms of the parameter $R$ that is the
%finite radius of a real solenoid \cite{kh1}.

Here we study the vacuum polarization of planar charged Dirac fermions
in a Coulomb potential. We express the induced charge
density in the vacuum via the exact Green's function, constructed
from solutions of the self-adjoint two-dimensional Dirac Hamiltonians with a (subcritical
and supercritical) Coulomb potential. To avoid misunderstanding, it should be noted that by Coulomb potential in 2+1 dimensions, we mean potential that decreases as $1/r$ with the distance from the source, having in mind that in a physical situation (e.g., in graphene), although the electrons move in a plane, their interaction with the external field of the Coulomb impurity occurs in a physical (three-dimensional) space and the electric field strength of the impurity is a three-dimensional (not two-dimensional) vector.
Therefore, the potential $A_0(r) \sim 1/r$ (and not $A_0(r) \sim \log r$,
as would be the case in 2+1 dimensions) does not satisfy the
two-dimensional Poisson equation with a pointlike source at the origin.

In order to see how the electron spin affects on the physical process under investigation, we also
consider a superposition of Coulomb and Aharonov--Bohm (AB) potentials. Then, the two-dimensional Dirac Hamiltonian will contain the term characterizing the interaction potential of the electron spin magnetic moment with AB magnetic field ${\bf H}=(0,\,0,\,H)=\nabla\times {\bf A}= B\pi\delta({\bf r})$ in the form $-seB\delta(r)/r$, which is singular and must influence the behavior of solutions at the origin. Here $s$ corresponds to the spin projection of a planar electron  on the $z$ (quantization) axis in three spatial dimensions.
We note that such kind of point interaction also appears in several Aharonov--Bohm-like problems \cite{khho0,asp1,sa1}.

We shall adopt the units where $c=\hbar=1$.

%\section{Induced vacuum charge density}

\section{Green's function for the self-adjoint two-dimensional Dirac Hamiltonians}

It is well known that the Dirac $\gamma^{\mu}$-matrix algebra is known to be represented in terms of the two-dimensional Pauli matrices  $\gamma^0= \sigma_3,\quad \gamma^1=is\sigma_1,\quad \gamma^2=i\sigma_2$ where  the parameter $s=\pm 1$ can be introduced to label two types of fermions in accordance with the signature of the two-dimensional Dirac matrices \cite{27}; for the case of massive fermions it can be applied to characterize two states of the fermion spin (spin "up" and "down")  \cite{4}.

The Dirac Hamiltonian for a fermion of the mass $m$ and charge
$e=-e_0<0$, which contains a parameter $s$ to describe the particle spin, in Coulomb ($A_0(r) =Ze_0/r\equiv a/e_0r$, $A_r=0$, $A_{\varphi}=0$, $a>0$) and   Aharonov--Bohm
($A_0=0$, $A_r=0$, $A_{\varphi}=B/r$) potentials ($r=\sqrt{x^2+y^2}$ and $\varphi=\arctan(y/x)$ are  circular cylindrical coordinates) is
\bb
 H_D=\sigma_1P_2-s\sigma_2P_1+\sigma_3 m-e_0A_0(r),\label{diham}
\ee
where $P_\mu = -i\partial_{\mu} - eA_{\mu}$ is the
generalized fermion momentum operator (a three-vector).
The Hamiltonian (\ref{diham}) should be defined as a self-adjoint operator in the Hilbert space
of square-integrable two-spinors $\Psi({\bf r})$.
The  total Dirac momentum operator  $J=-i\partial/\partial\varphi+ s\sigma_3/2$ commutes with  $H_D$.
Eigenfunctions of the Hamiltonian (\ref{diham}) are (see, \cite{khho,khlee1})
\bb
 \Psi(t,{\bf r}) = \frac{1}{\sqrt{2\pi r}}
\left( \begin{array}{c}
f(r)\\
g(r)e^{is\varphi}
\end{array}\right)\exp(-iEt+il\varphi)~, \label{three}
\ee
where $E$ is  the fermion energy, $l$ is the integer quantum number.
The wave function $\Psi$ is an eigenfunction of the
operator $J$ with eigenvalue $j=\pm (l+s/2)$ in terms of the angular momentum $l$ and
\bb \check h F(r)= EF(r), \quad F(r)=\left(
\begin{array}{c}
f(r)\\
g(r)\end{array}\right), \label{radh}\ee
where
\bb
\check h=is\sigma_2\frac{d}{dr}+\sigma_1\frac{l+\mu+s/2}{r}+\sigma_3m-\frac{a}{r},\quad \mu\equiv e_0B. \label{radh0}
\ee

The induced current density due to to vacuum polarization is
determined by the three-vector $j_{\mu}({\bf r})$, which is
expressed via the single-particle Green function of the Dirac equation as
\bb
 j_{\mu}({\bf r})=-\frac{e}{2}\int\limits_{C}\frac{dE}{2\pi i}{\rm tr}G({\bf r}, {\bf r'}; E)|_{{\bf r}={\bf r'}}\gamma_{\mu},
\label{cur0}
\ee
where $C$ is the path in the complex plane of $E$ enclosing all the singularities
along the real axis $E$ depending upon the choice of the Fermi level $E_F$.
The Green's function $G$ can be expanded in eigenfunctions of the operator $J$.
The radial parts (the doublets) of above eigenvalues must satisfy  the two-dimensional Dirac equation ({\ref{radh}).
Then the radial partial Green's function $G_l(r, r'; E)$ is given by (just as in 3+1 dimensions \cite{grrein})
\bb
G_l(r, r'; E)\gamma^0=\frac{1}{{\rm W}(E)}[\Theta(r'-r)U_R(r)U^{\dagger}_I(r')+
\Theta(r-r')U_I(r)U^{\dagger}_R(r')],
\label{green5}
\ee
where ${\rm W}(E)$ is the  ($r$-independent) Wronskian, defined by two doublets $V$ and $F$
as ${\rm Wr}(V, F)=Vi\sigma_2F=(v_1f_2-f_1v_2)$, where indexes denote upper and lower doublet components, and $U_R(r)$ and $U_I(r)$ are the regular and irregular solutions of the radial Dirac equation $(\check h-E)U(r)=0$; the regular (irregular) solutions are integrable at $r\to 0$ ($r\to\infty$). We see that the problem is reduced to constructing the self-adjoint radial Hamiltonian $\check h$  in the Hilbert space of  doublets $F(r)$ square-integrable on the half-line.

Since the initial radial Dirac operator  is not determined as an unique self-adjoint
operator the additional specification of its domain, given with the real parameter $\xi$ (the self-adjoint extension parameter)  is required in terms of the self-adjoint boundary conditions.
Any correct doublet $F(r)$   of the Hilbert space
 must satisfy the self-adjoint boundary condition \cite{17,khlee0,khlee}
\bb
 (F^{\dagger}(r)i\sigma_2 F(r))|_{r=0}= (\bar f_1f_2-\bar f_2f_1)|_{r=0} =0. \label{bouncon}
\ee
Physically, the self-adjoint boundary conditions  show that the probability current density  is equal to zero at the origin.

We shall apply as the solutions of the radial Dirac
equation (\ref{radh0}) the doublets represented in the form
\bb
F_R = \left( \begin{array}{c} f_R(r, \gamma, E) \\ g_R(r, \gamma, E) \end{array} \right),
F_I = \left( \begin{array}{c} f_I(r, \gamma, E) \\ g_I(r, \gamma, E) \end{array} \right),
\label{WF1}
\ee
where
\bb
f_R(r, \gamma, E)=\frac{\sqrt{m+E}}{x} \left( A_R M_{aE/\lambda +s/2, \gamma}
(x) + C_R M_{aE/\lambda -s/2, \gamma} (x) \right), \phantom{mmmmmmmmmmm}\nonumber \\
g_R(r, \gamma, E)=\frac{\sqrt{m-E}}{x} \left( A_R M_{aE/\lambda +s/2, \gamma}
(x) - C_R M_{aE/\lambda -s/2, \gamma} (x) \right),\quad \frac{C_R}{A_R}=\frac{s\gamma -aE/\lambda}{\nu+ma/\lambda},
\label{base}
\ee
\bb
f_I(r, \gamma, E)=\frac{\sqrt{m+E}}{x} \left( A_I W_{aE/\lambda +s/2, \gamma}
(x) + C_I W_{aE/\lambda -s/2, \gamma} (x) \right),\phantom{mmmmmmmmmmm} \nonumber \\
g_I(r, \gamma, E)=\frac{\sqrt{m-E}}{x} \left( A_I W_{aE/\lambda +s/2, \gamma}
(x) - C_I W_{aE/\lambda -s/2, \gamma} (x) \right),\quad
\frac
{C_I}{A_I}=(ma/\lambda-s\nu)^s. \label{BASE}
\ee
Here
\bb
x=2\lambda r, \quad \lambda = \sqrt{m^2 - E^2}, \quad \gamma = \sqrt{\nu^2
-a^2}, \quad \nu=|l+\mu+s/2|, \label{Note}
\ee
$A_R, A_I, C_R, C_I$ are numerical coefficients and the Whittaker functions $M_{a,b}(x)$ and $W_{c,d}(x)$ represent the regular and irregular solutions.

For $a^2\leq \nu^2$ $\gamma$ is real, for $a^2>\nu^2$
$\gamma=i\sqrt{a^2-\nu^2}\equiv i\sigma$ is imaginary.
The quantities $q=\sqrt{\nu^2-\gamma^2}$ and  $q_c=\nu \Leftrightarrow\gamma=0$ are called the effective and   critical charge, respectively; it is helpful also to determine $q_u=\sqrt{\nu^2-1/4}\Leftrightarrow\gamma=1/2$.
We note that all the fermion states are doubly degenerate with respect
to the spin parameter $s$ at $\mu=0$.

In the subcritical range,  for $q\leq q_u$ ($\gamma\geq 1/2$),   only solutions $F_R(r)$ vanishing at $r=0$ can be chosen as the regular ones; they satisfy (\ref{bouncon}). For $q_u<q<q_c$  ($0<\gamma<1/2$), the regular solutions $U_R(r)$ satisfying the self-adjoint boundary condition (\ref{bouncon}) should be chosen in the form of linear combination of  $F_R(r)$ and $F_I(r)$ \cite{17,khlee1}
\bb
U_R(r)=F_R(r)+\xi F_I(r)
\label{mainf}
\ee
and the Wronskian is
\bb
{\rm Wr}(F_R, F_I)\equiv {\rm W}(E, \gamma) = (g_R f_I - f_R g_I) = -2 A_R A_I \frac{\Gamma
(2\gamma)}{\Gamma (\gamma + 1/2 -s/2 -aE/\lambda)}
\frac{s\gamma}{\nu+ma/\lambda}
\label{wr1}
\ee
where $\Gamma(z)$ is the Gamma function \cite{GR} and, therefore, in the subcritical range
the single-particle Green function is completely determined.
One can show that the  contribution into the induced charge density coming for
$0<\gamma<1/2$ is small at any $\xi$, therefore it is enough to consider the case $\xi=0$.
Thus, we can chose as the regular solutions  the functions $F_R(r)$ for all $\gamma>0$
\bb
{\rm tr} G_{\nu} ({\bf r}, {\bf r'}; E)|_{{\bf r}={\bf r'}} \gamma^0 = \sum_{s={\pm 1}}
\sum^{\infty}_{l=-\infty}\frac{f_I f_R + g_I g_R}{2\pi s{\rm W}(E, \gamma)}.
\label{TRACE}
\ee
Performing some simple calculations, we obtain
\bb
{\rm tr} G_{\nu} ({\bf r}, {\bf r'}; E)|_{{\bf r}={\bf r'}} \gamma^0 = -\frac{1}{2\pi \lambda^2 r^2}
\sum^{\infty}_{l=-\infty} \frac{\Gamma(\gamma -
aE/\lambda)}{\Gamma(2\gamma +1)} \left[ (m^2 a/\lambda +E(x
-2aE/\lambda -1 )) M_{aE/\lambda +1/2, \gamma} (x) W_{aE/\lambda
+1/2, \gamma} (x) + \right. \nonumber \\
\left. + m^2a [(\gamma -aE/\lambda)/\lambda] M_{aE/\lambda -1/2,
\gamma} (x) W_{aE/\lambda -1/2, \gamma} (x) +
Ex\frac{d}{dx}(M_{aE/\lambda +1/2, \gamma} (x) W_{aE/\lambda +1/2,
\gamma} (x)) \right]. \phantom{mmmm}\label{Calcul1}
\ee

We note that the singularities of $G_{\nu}({\bf r}, {\bf r'}; E)$ are
simple poles associated with the discrete spectrum for $-m<E<m$,
and two cuts ($(-\infty,-m]$ and $[m,\infty)$) associated with the continuous
spectrum for $|E|\geq m$ \cite{20}.

  The path $C$ in Eq. (\ref{cur0}) may be
deformed to run along the singularities on the real $E$ axis as follows:
$C=C_-+C_p+C_+$, where $C_-$ is the path along the negative
real $E$ axis (${\rm Re}E<0$) from $-\infty$ to $0$ turning
around at $E=0$ with positive
orientation, $C_p$ is a circle around the bound states singularities
with $-m< E<0$ (if we chose $E_F=-m$),
and $C_+$ is the path along the positive real $E$ axis (${\rm Re}E>0$)
from $\infty$ to $0$ but with negative orientation (i.e. clockwise path)
turning around at $E=0$ \cite{grrein,kh1}.

The contour of integration $C$ with respect to $E$ can be deformed to coincide
with the imaginary axis and we obtain as a result:
\bb
j_0({\bf r}) = -e \int\limits_{-\infty}^{\infty}\frac{dE}{2\pi}{\rm tr}
G_{\nu} ({\bf r},{\bf r}, iE) \gamma^0. \label{Density}
\ee
Let us represent  $\mu=[\mu]+\alpha\equiv n+\alpha$, where
$n=0, 1, 2, \ldots$ for $\mu>0$, $n=-1, -2, -3, \ldots$ for $\mu<0$ and $1>\alpha\ge 0$;
denote  $\nu_{\pm} = l \pm \alpha +1/2,\quad \gamma_{\pm} =\sqrt{\nu^{2}_{\pm} -a^2}$,
where here and in all formulas below $l\equiv l+n$. Since signs of $e$ and $B$ are
fixed it is enough to consider only the case $\mu>0$.
Then, by means of formula \cite{GR}
\bb
M_{aE/\lambda \pm 1/2,\gamma}(x) W_{aE/\lambda \pm 1/2,\gamma}(x)
=\frac{x\Gamma(2\gamma +1)}{\Gamma(1/2 +\gamma -aE/\lambda \mp 1/2)}
\int\limits_{0}^{\infty} e^{-x\cosh s}
[\coth(s/2)]^{2aE/\lambda \pm 1} I_{2\gamma}(x\sinh s) ds, \phantom{mmm}
\label{Form}
\ee
after long calculations, we represent the induced charge density in the form
\bb
j_0(r) = - \frac{2e}{\pi^2 r } \sum_{l=0}^{\infty}
\int\limits_{0}^{\infty} dE \int\limits_{0}^{\infty} dt
e^{-2\lambda r \coth t} \left( 2a \cos(2aE/\lambda) \coth t (I_{2\gamma_{+}}
(2\lambda r/\sinh t) + I_{2\gamma_{-}} (2\lambda
r/\sinh t)) -\right. \nonumber \\
- \left. \frac{2Er}{\sinh t} \sin(2aE/\lambda)
(I^{\prime}_{2\gamma_{+}} (2\lambda r/\sinh
t)+I^{\prime}_{2\gamma_{-}} (2\lambda r/\sinh t)) \right).
\label{Res1}
\ee
where $\lambda =\sqrt{m^2+E^2}$, $I_{\mu}(z)$ is the modified Bessel function of the first kind  and the prime (here and below) denotes the derivative of function with respect to argument.
 We note that $j_0$ is odd with respect to charge $e$.

\section{Renormalized induced vacuum charge}

Due to the  mass $m$ the renormalization of Eq. (\ref{Res1}) can be performed
as well as in the massive quantum electrodynamics and   it also is convenient to do
the renormalization in momentum space:
\bb
j_0(z)\equiv \rho(z) =\int d{\bf r} e^{i{\bf q\cdot r}} j_{0} (r)
=\frac{2e}{\pi} \sum_{l=0}^{\infty} \int\limits^{\infty}_{0} dx
\int\limits^{\infty}_{0} dt \int\limits^{\infty}_{0} dy \frac{\sinh
t}{2b} e^{-y \cosh t} J_{0} (zy \sinh t/2b) f(y,t), \nonumber \\
f(y,t)=\frac{xy}{b} \sin(ct) (I^{\prime}_{2\gamma_{+}} (y)
+I^{\prime}_{2\gamma_{-}} (y)) - 2a \coth t \cos(ct)
(I_{2\gamma_{+}} (y) +I_{2\gamma_{-}} (y)). \label{Imp}
\ee
Here  $z=|q|/m$, $x=E/m$, $b=\sqrt{1+x^2}$, $y=2bmr/\sinh t$, $c=2ax/b$.

We can satisfy the obvious physical requirement
of vanishing of the total induced charge,
because the induced charge density diminishes rapidly at distances $r\gg 1/m$.
Since the presence of external fields do not give rise to additional divergences
in expressions of perturbation theory
it is enough (and convenient) to carry out the renormalization  in the subcritical range.
Then, calculations which we need to perform are similar to those described
in \cite{01,02,12,ms0}. We introduce the renormalized induced charge  in momentum representation as $\tilde\rho(z)=\lim_{\Lambda\to \infty}[\rho(z)-\lim_{z\to 0}\rho(z)]$
with an ultraviolet cutoff $|E|<\Lambda$.  Because the mass $m$ is the only
dimensionful parameter in the Green function  the resulting dimensionless function
$\tilde\rho(z)$ can depend only on the ratio $q/m$.

As $a<1/2$, the terms of different order in $a$ behave differently.
We can see it in terms of perturbation theory. Indeed, the
linear in $a$ term  corresponds to the diagram of the polarization operator
in the one-loop approximation  and its renormalization  coincides with
the usual procedure of renormalizing the polarization operator.
The terms proportional to $a^3$ correspond to
diagrams of the type of photon scattering by photon and, in difference on the
case of the 3D quantum electrodynamics (see \cite{01,02,03,ms0}) they are finite. However their
regularization must still be carried out in the considered case  due to the requirements of gauge invariance, which, in particular,  determine the behaviors of the scattering amplitude at small  $|q|/m$.

{\bf Massless case}.
We shall first consider  the more complicated case with $m=0$.
It will be noted that the massless fermions do not have spin degree of freedom in 2+1 dimensions \cite{jacknai}, nevertheless, solutions of the Dirac equation for massless fermions in the superposition of Coulomb and AB potentials keep the introduced spin parameter.
The leading term of the induced charge density
at the limit $m\to 0$ (or $z\to \infty$) is a constant $Q_{ind}=\lim_{m\to 0}\tilde\rho(z)$. So  $Q_{ind}$ is the induced charge density localized in
the point ${\bf r}=0$ and, therefore, the total induced charge density in
coordinate space has the form
 \bb \tilde\rho({\bf r})=Q_{ind}\delta({\bf
r})+\rho_{dist}({\bf r}), \label{qtot} \ee where $\rho_{dist}({\bf
r})$ is the so-called distributed charge density. The induced and distributed
charge densities have opposite signs. The total distributed charge
$\int\rho_{dist}({\bf r})d{\bf r}$ is equal to $-Q_{ind}$.

We have carried out long calculations and got the renormalized induced charge in the subcritical regime  that is exact in the parameter $a$:
\bb
Q_{ind} = Q_1(e_0a,\alpha) + Q_r(e_0a,\alpha). \label{Tot}
\ee
Here
\bb Q_1(e_0a,\alpha)= \frac{2ea}{\pi}
\sum_{l=0}^{\infty} \left( (l+1/2+\alpha) \psi^{\prime}
(l+1/2+\alpha) +(l+1/2-\alpha) \psi^{\prime} (l+1/2 - \alpha) - 2
- \frac{l+1/2}{(l+1/2)^2 - \alpha^2}
\right), \label{Q1}
\ee
\bb
Q_r(e_0a,\alpha) = \frac{2e}{\pi} \sum^{\infty}_{l=0}{\rm Im} \left[ \ln (\Gamma
(\gamma_{+} - ia)\Gamma (\gamma_{-} - ia)) + \frac{1}{2} \ln (
(\gamma_{+} - ia)(\gamma_{-} - ia)) - \right. \nonumber \\
- ((\gamma_{+} -ia) \psi (\gamma_{+} -ia) + (\gamma_{-} -ia) \psi
(\gamma_{-} - ia)) + ia \frac{l+1/2}{(l+1/2)^2 - \alpha^2} -
\nonumber \\
\left. - ia ((l+1/2 +\alpha) \psi^{\prime} (l+1/2 +\alpha) +
(l+1/2 - \alpha) \psi^{\prime} (l+1/2 - \alpha)) \right],
\label{Qr}
\ee
where $\psi(z)$ is the logarithmic derivative of Gamma function \cite{GR}.
The expression (\ref{Tot})  is exact in $a$ in the subcritical range and $\alpha$, is odd (even) with respect to $a$ ($\alpha$). It  is in agreement  with result obtained in \cite{12} for $\alpha=0$;
the coefficient of the $a^3$ term at $\alpha=0$ was also found in perturbation
theory \cite{bsaso}. The induced vacuum charge $Q_{ind}$ is negative; thus it has a
screening sign, leading to a decrease of the effective charge of Coulomb center.
 For $\alpha \ll 1$,  we find
\bb
Q_1(e_0a, \alpha) = ea\pi/4 + ea\pi (2 \ln2 + 1 -\pi^2/4) \alpha^2  \approx ea\pi(0.25 -0.04\alpha^2).
 \label{ApprQ}
\ee
This expression reflects the linear one-loop polarization contribution. The first term in  Eq. (\ref{ApprQ}) coincides  with result obtained in \cite{12,as11b,bsaso}. We  note that the contribution into $Q_1(e_0a,\alpha)$ from AB potential arises in the presence of Coulomb field only,  is small
and has opposite sign compared with a pure Coulomb one.

Since Eq. (\ref{Tot}) is even with respect to $\mu$, it is clear that
the fermion spin does not contribute in the induced vacuum charge in the subcritical regime.
Moreover, such a (spin) problem has a physical meaning  in the subcritical range only, when
the expression for induced vacuum charge can be expanded in ascending power series of $a, \alpha$.

In the supercritical range, we shall consider the only Coulomb problem  putting $\alpha=0$ in $\gamma, \nu$.
Now $\gamma=i\sigma$, therefore, we need directly to determine
the Green's function specified by boundary conditions (\ref{bouncon}).
We straightforward construct the Green function in the form
(\ref{green5}) in which the regular solutions $U_R(r)$,
satisfying (\ref{bouncon}), have to be chosen in the form of
linear combination of the functions $F_R(r)$ and $F_I(r)$.
For this range,  the above two solutions $F_R(r)$ and $F_I(r)$ become
oscillatory  with the imaginary exponent and  it is convenient to use in this range
the self-adjoint extension parameter  $\theta$ \cite{khlee1,20}, related to $\xi$ by
\bb
\frac{A_R}{\xi A_I} = e^{2i\theta} \left(\frac{2\lambda}{E_0}\right)^{-2i\sigma}
\frac{\nu +a(m+E)/\lambda +is\sigma}{\nu +a(m+E)/\lambda
-is\sigma} \frac{\Gamma(2i\sigma)}{\Gamma(1/2-s/2-aE/\lambda
+i\sigma)} - \frac{\Gamma(-2i\sigma)}{\Gamma(1/2-s/2-aE/\lambda
-i\sigma)},\label{xithet}
\ee
where $\pi\geq \theta\geq 0$ and a positive constant $E_0$ gives an energy scale.

The Green's function has a discontinuity, which is solely associated with the appearance
of its singularities situated on a second (unphysical) sheet ${\rm Re}E<0, {\rm Im}E<0$
of the complex plane $E$ at $q>q_c$; these
singularities are determined by complex roots of equation ${\rm W}(E, i\sigma)=0$ and describe
the infinite (for massless fermions) number of quasistationary  states with complex ''energies'' $E=|E|e^{i\tau}$ with $\tau>\pi$; for $\sigma\ll 1$  their energy spectrum was found in \cite{20}:
\begin{align}
 E_{k,\theta}\equiv {\rm Re}E = E_0 \cos(\tau)
 \exp(-k/2\sigma+\theta/\sigma+\pi\coth\pi a/2a),
\label{energyr}
\end{align}
where $\tau \approx 1/2a+\mathrm{Im}\psi(ia)+\pi/2$.  These quasi-localized resonances have negative energies, thus they are situated in the hole sector.
For $\sigma\ll 1$ the imaginary part
${\rm Im}E= \tan\tau E_{k,\theta}\ll {\rm Re} E$ is very small and, therefore, the resonances are practically stationary states \cite{20}.
Physically, the self-adjoint extension parameter can be
interpreted  in terms of the cutoff radius $R$ of a Coulomb potential. For this, for example, we can compare  Eq. (\ref{energyr}) with the spectrum
of supercritical resonances in the cutoff Coulomb potential \cite{as11b,ggg} and
 approximately derive $\theta \sim \sigma[c(a)+ \ln E_0R]$, where $c(a)$ does not depend on $R$.
We  note that  the cutoff radius $R$ rather relates to a renormalized critical coupling
that is also characterized by a logarithmic singularity at $mR\ll 1$ in massive case \cite{ggg,khho}.

The simplest way to include these resonances in the induced charge density is
to carry out the integral in $E$ from $-\infty$ to $0$ along the path $S$
taking into account the singularities on the second sheet.
After some calculations, we represent the induced charge (electron) density (\ref{cur0}) as
the sum of contributions from the subcritical and supercritical ranges, which have to be treated separately
\bb
j_0(r) = -e \int\frac{dE}{8\pi^2 i}{\rm tr}
G_{\nu} (r, r, E) \gamma^0 =-e\int\limits_{C}\frac{dE}{8\pi^2
i}\sum_{s=\pm 1} \sum^{\infty}_{l=-\infty}\frac{f_I(r, \gamma, E) f_R(r, \gamma, E) + g_I(r, \gamma, E)g_R(r, \gamma, E)}{s{\rm W}(E, \gamma)}-\nonumber \\ -e\int\limits_{S}\frac{dE}{8\pi^2
i}\sum_{l,s: \nu<a} \frac{\xi (f_I^2(r, i\sigma, E) + g_I^2(r, i\sigma, E))}{s{\rm W}(E, i\sigma)}
=Q_{ind}(r) +j_{supcr} (r).\phantom{mmmmmm} \label{Den}
\ee
For the supercritical range  $\gamma=i\sigma$, $0\geq \theta\geq \pi$, the sum in second term $j_{supcr}$ is taken  over $l$ of  $a^2>(l+s/2)^2$.
Then the paths $C, S$ can be deformed to coincide with the imaginary axis $E$.

The first term ($Q_{ind}(r)$) in Eq. (\ref{Den}) was calculated and explicitly represented by Eqs. (\ref{qtot}) and (\ref{Tot}). The second term is convergent and  its contribution to
the induced charge density can be directly evaluated at $m=0$. Having performed simple calculations
we leads $j_{supcr}$ to
\bb
j_{supcr}(r) =\frac{e}{8\pi^2 r^2}\sum_{l,s: \nu<a}
\frac{s\nu^{s+1}}{\sigma \Gamma(2i\sigma) \Gamma(-2i\sigma)}
\int\limits_{-\infty}^{0} \frac{dE}{E\omega(\sigma)} \Gamma(i\sigma +(1-s)/2 -iaE/|E|)\times \nonumber \\
\times \Gamma(-i\sigma
+(1-s)/2 -iaE/|E|) W_{iaE/|E| +s/2, i\sigma} (2|E|r)W_{iaE/|E| -s/2, i\sigma}
(2|E|r), \label{supcr1}
\ee
where
\bb
\omega(\sigma) = 1- e^{2i\theta} \left(
\frac{2|E|}{E_0}\right)^{-2i\sigma} \frac{\nu+iaE/|E|
+is\sigma}{\nu+iaE/|E| -is\sigma}
\frac{\Gamma(2i\sigma)}{\Gamma(-2i\sigma)}\frac {\Gamma(-i\sigma
+(1-s)/2 -iaE/|E|)} {\Gamma(i\sigma +(1-s)/2 -iaE/|E|)}.
\label{omeg1}
\ee

Rewrite  $(2|E|/E_0)^{-2i\sigma}$ as $\exp(-2i\sigma \ln (|E|/E_0))$. As far as the integrand (\ref{supcr1}) decreases exponentially at $|E|\gg 1/r$ and strongly oscillate at $|E|\to 0$,
the main contribution to the integral over $E$  is given by the region $|E|\sim 1/r$.
So in order to evaluate $j_{supcr}$ we replace $|E|$ by $1/r$ in the log-periodic term of the integrand
(\ref{omeg1}) and  obtain
\bb
j_{supcr}(r) = -\frac{e}{8\pi^2 r^2}\sum_{l,s: \nu<a}
\frac{s\nu^{s+1}\Gamma(i\sigma
+(1-s)/2 +ia)}{\sigma\omega_-(\sigma) \Gamma(2i\sigma) \Gamma(-2i\sigma)}
\Gamma(-i\sigma +(1-s)/2 +ia) \times \nonumber \\
\times \int\limits_{0}^{\infty} \frac{dE}{E}W_{-ia +s/2,
i\sigma}(2Er)W_{-ia -s/2, i\sigma} (2Er), \phantom{mmmmmmmmmm} \label{supcr2}
\ee
where
\bb
\omega_-(\sigma) = 1- e^{2i\theta +2i\sigma \ln(E_0 r)}
\frac{\nu - ia +is\sigma}{\nu - ia -is\sigma}
\frac{\Gamma(2i\sigma)}{\Gamma(-2i\sigma)}\frac {\Gamma(-i\sigma
+(1-s)/2 + ia)} {\Gamma(i\sigma +(1-s)/2 + ia)}. \label{omega1}
\ee
Because of the complex singularities on the unphysical sheet at $q>q_c$,
the Green's function and  $j_{supcr}(r)$ are complex though for $\sigma\ll 1$ their imaginary parts
are small.  In terms of the physics the complex Green's function probably reflects the lack of
stability of chosen (for constructing Green function) neutral vacuum
for $q>q_c$ (see, also \cite{grrein}).

Now we can integrate  in Eq. (\ref{supcr2}) using formula \cite{GR}
\bb
\int\limits_{0}^{\infty} \frac{dE}{E}W_{-ia +s/2,
i\sigma}(2Er)W_{-ia -s/2, i\sigma} (2Er) = \frac{\pi}{s \sin(2\pi i\sigma)}\times \phantom{mmmmmmmmmm} \nonumber
\\ \times \left[
\frac{1}{\Gamma((1-s)/2 +ia +i\sigma) \Gamma((1+s)/2 +ia
-i\sigma)} -  \frac{1}{\Gamma((1-s)/2
+ia -i\sigma) \Gamma((1+s)/2 +ia +i\sigma)} \right]
 \label{F1}
\ee
and after simple transformations we  finally find the induced charge density in the supercritical range as
\bb
j_{supcr}^r(r) = \frac{e}{2\pi^2  r^2}\sum_{l,s: \nu<a}{\rm Re}\frac{\sigma}{\omega_-(\sigma)}. \label{sup4}
\ee

The main effect, arising at the supercritical regime, is that the induced
vacuum polarization for noninteracting massless fermions has a power law form ($\sim c/r^2$)  whose
coefficient is log-periodic functions with respect
to the distance from the origin. In the subcritical regime the induced vacuum
charge  is localized at origin and exhibits no long range tail.
As an example, we consider Eq. (\ref{sup4}) for $1/2<a<3/2$, when just the lowest $l=-1, 0$ channels are supercritical, and find
\bb
j_{supcr}^r(r) = \frac{e}{\pi^2  r^2}\sigma_0{\rm Re}\frac{2-|A|ze^{2i\theta +2i\sigma_0 \ln(E_0 r)+i\psi}}{1-|A|ze^{2i\theta +2i\sigma_0 \ln(E_0 r)+i\psi}+|A|^2[(a-\sigma_0)/(a+\sigma_0)]e^{4i\theta +4i\sigma_0 \ln(E_0 r)+2i\psi}}, \label{fsup}
\ee
where
$$
 A=\frac{\Gamma(2i\sigma_0)\Gamma(-i\sigma_0+ia)}{\Gamma(-2i\sigma_0)\Gamma(i\sigma_0+ia)},\quad
 z=2\frac{a-\sigma_0}{a},\quad \sigma_0=\sqrt{a^2-1/4},
$$
$$
\psi \equiv {\rm Arg} A = -\pi-2{\cal C}\sigma_0 +\sum\limits_{n=1}^{\infty}\left(\frac{2\sigma_0}{n}-2\arctan\frac{2\sigma_0}{n}
+\arctan\frac{2n\sigma_0}{n^2+1/4}\right).
$$
Here ${\cal C}=0.57721$ is Euler's constant.
For small $\sigma_0\ll 1$, Eq. (\ref{fsup}) takes the simplest form
\bb
j_{supcr}^r(r) = \frac{e\sigma_0}{\pi^2  r^2}. \label{1fsup}
\ee

It is of importance that the induced charge density $j_{supcr}^r(r)$ (\ref{1fsup})  at $\sigma_0\ll 1$ does not contain at all the self-adjoint extension parameter $\theta$. From the physical point of view, when  the Coulomb center charge is suddenly increased from subcritical
to supercritical values,  the transition will occur from the subcritical range to the supercritical one,  and then a small change in $q$ such that $q>q_c$ leads to a sudden change in the character of a physical phenomenon due to emerging of infinitely many resonances with negative energies.
However, the character of a physical phenomenon itself must be due only to physical  (but not mathematical)  reasons.
We also note that the expression (\ref{1fsup}) is in agreement  with results
obtained in \cite{as11b} for the problem of vacuum polarization of supercritical
impurities in graphene by means of scattering phase analysis.

For large $a\gg 1, \sigma\approx a-l^2/2a$, the induced charge density can approximately
be represented as
\bb
{\rm Re}j_{supcr}(r) = \frac{e}{\pi^2 r^2}\sum_{l<a}\sqrt{a^2-l^2},\quad {\rm Im}j_{supcr}(r) =0. \label{im0sup}
\ee

\section{Screening of Coulomb center charge}

The expressions for induced charge densities, found for noninteracting fermions, can be
used to describe screening of the Coulomb center charge in an interacting fermion system.
Notice that the induced charge has a screening sign, leading to a decrease
of the effective Coulomb center charge.
The strongly localized distribution of the induced charge
in the subcritical regime implies that the Coulomb charge merely renormalizes the
strength of the Coulomb center leading to the replacement $a\to a_{eff}$  in the Coulomb potential
where $a_{eff}$ are real solutions  of corresponding equation taking
into account also electron-electron interactions in the Hartree approximation \cite{12}
\bb
a_{eff}= a-e[Q_1(e_0a_{eff})+Q_r(e_0a_{eff})].
\label{subren}
\ee
It  is essential  that Eq. (\ref{subren}) do not have
solutions with $a_{eff}\geq 1/2$ for $a=e_0^2$, which means that the effective
Coulomb center charge remains subcritical \cite{12}.
Nevertheless, for Coulomb center charge $2e_0, 3e_0$ and
higher, the effective charge can become supercritical at certain values
of $e_0^2$ \cite{12}.

In the the supercritical regime, the induced charge density (\ref{fsup}) causes a modification of Coulomb potential at large distances and since the infinite number of quasistationary
states emerge they should contribute significantly to shield the Coulomb center.
It is clear that at least for small $\sigma$
a planar electron at some distance $r$ from the Coulomb center feels the effect
of an effective point charge consisting of the ``bare" charge of the Coulomb center  subtracted
from the induced screening charge within the annulus $r_0, r,\quad r_0<r$.
At first we can found the induced charge within the annulus
$Q$  integrating Eq. (\ref{1fsup})
\bb
Q(r)=-2\frac{e_0 \sigma_0}{\pi}\ln\frac{r}{r_0}
\label{charge0}
\ee
and then treat $Q(r)$  as an effective point charge.

Since the logarithmic  term represents the renormalization of the (supercritical)
charge of Coulomb center, we can write instead of  Eq. (\ref{charge0}) a differential
equation, which defines a self-consistent renormalization of the effective coupling
$g\equiv a_{eff}$  like the differential
equation of the renormalization group  (see, for instance,  \cite{as11b,review}):
\bb
\frac{dg}{d\ln(r/r_0)}=-2\frac{e_0^2 \sigma_0}{\pi}.
\label{renorm}
\ee
In this way, it can be seen
that the effective coupling $g$  will tend to the constant $1/2$  within a finite distance
$$
r=r_0e^{-(2\pi/e_0^2)\ln[2g+\sqrt{4g^2-1}]}
$$
from the Coulomb center.

The renormalization group treatment is applicable when
the right-hand side of Eq. (\ref{renorm}) is small, i.e. for $\sigma_0\ll 1$.
Therefore for $\sigma_0\ll 1$,  the vacuum of planar charged electrons with Coulomb potential
self-consistently rearranges itself so that electrons at distances $r>r_0$ never
feel a supercritical effective coupling irrespective of the ``bare" supercritical Coulomb
center charge (see, \cite{as11b,review}).

It is well to note that in the convenient quantum electrodynamics
the vacuum polarization charge in super-heavy
nuclei behaves in such a way as to reduce the supercritical charge
of nucleus to the threshold value \cite{muraf} (see also \cite{terkh},
where the problem was investigated for super-heavy
nuclei in the presence of a superstrong constant uniform magnetic field).

\section{Vacuum polarization of planar charged massive fermions}

We now briefly  address to the vacuum polarization induced by the Coulomb potential
in massive case. If the  Coulomb center charge is subcritical the massive case has a well defined infinite spectrum of bound solutions situated on the physical sheet, which for $\gamma\geq 1/2, a<1/2, \xi=0$ is \cite{khho}
 \bb
 E_{k,l} = m\frac{k +  \sqrt{\nu^2-a^2}}
 {\sqrt{[k + \sqrt{\nu^2-a^2}]^2+a^2}}, \quad \nu =l+1/2;\quad k, l= 0, 1, 2 \ldots ,
\label{spectrumw}
\ee
We see that all the energy levels are doubly degenerate with respect to  $s$.
It can be easily shown  that
the spectrum accumulates at the point $E = m$, and its asymptotic form as
$n=k+l\to \infty$ is given by the nonrelativistic formula
$\epsilon_n=m-E_n=ma^2/n^2$. The problem of finding the spectra of self-adjoint extensions of the radial Hamiltonian in the Coulomb and Aaronov-Bohm potentials
in 2+1 dimensions was solved in \cite{khlee0} where, in particular, it was shown that
the spectrum accumulates at the point $E = m$ and is described by the same asymptotic
formula (without AB potential), independent of $\xi$, i.e. $\epsilon_n=m-E_{n,\xi}=ma^2/n^2$.

In the massive case the vacuum polarization of planar charged fermions manifests itself by modifying the Coulomb potential. Therefore, it is rewarding to calculate the polarization corrections to the Coulomb potential. As applied to the vacuum polarization we shall assume that none of the bound levels are occupied. If $a\ll 1$ we can estimate these polarization corrections in the first order in $a$.
For three spatial dimensions, the potential taking into account the polarization corrections of the first order in $a$ to the Coulomb potential is the Uehling-Serber potential.
In terms of perturbation theory, these corrections correspond to the polarization operator
in the lowest order in interaction. Performing the integrations and summation in Eq. (\ref{Imp}) with taking  only the linear in $a$ terms into account,  for the renormalized induced
Coulomb center charge, we obtain
\bb
Q_m(|q|)=-\frac{a}{e_0}\frac{\Pi(-{\bf q}^2)}{|{\bf q}|},
\label{po}
\ee
where, as it should be,
$$
\Pi(-{\bf q}^2)=\frac{e_0^2}{8\pi}\left(\frac{4m^2-{\bf q}^2}{\sqrt{{\bf q}^2}}\arctan\sqrt{\frac{{\bf q}^2}{4m^2}}-2m\right)
$$
is the polarization operator in the first order of perturbation theory.
After some transformations the induced charge distribution $Q_m(r)\equiv a_{eff}^m/e_0$
(here $a_{eff}^m$ is the effective coupling) takes the form in the coordinate space:
\bb
Q_m(r)=-e_0a\int\limits_{1}^{\infty}\frac{dx}{x^3\sqrt{x^2-1}}e^{-2mrx}.
\label{q1c}
\ee
The integral is calculated in limits $mr\ll 1$ and $mr\gg 1$ and as a result we find
\bb
Q_m(r)\approx -   e_0a\left[\frac{\pi}{4}-Cmr\right], \quad mr\ll 1, 1\gg C\gg mr,
\label{smalmr}
\ee
where the first term on the right of Eq. (\ref{smalmr}) was already calculated (see Eq. (\ref{Q1})), and
\bb
Q_m(r)\approx -e_0a\sqrt{\frac{4\pi}{mr}}e^{-2mr},  \quad mr\gg 1.
\label{bigmr}
\ee
We see that even at small distances from the Coulomb center, the finite mass contribution to the
induced vacuum charge is small and insignificantly distorts the Coulomb potential only at distances of the Compton length $r\sim 1/m$. The  induced charge   has a screening sign.

In the supercritical regime the finite mass contribution to the vacuum polarization easier to
estimate,  at least when $\sigma_0\equiv \sqrt{a^2-1/4}\ll 1$.
Indeed, if the Coulomb potential charge is suddenly increased from subcritical
to supercritical values then the only  lowest energy level  dives into the
negative energy continuum and becomes a resonance with ``complex energy''$E=|E|e^{i\tau}$.
 There appears the pole on the unphysical sheet $\tau>\pi$, counted now as a ``hole'' state.
Using results of Ref. \cite{khlee0}, one can show  the energy of dived state
${\rm Re}E=-(m+\epsilon), \epsilon\to +0$,  is determined by the following transcendental equation
\bb
 \arg\Gamma(2i\sigma_0)-\sigma_0{\rm Re}\psi(-iz)-(\sigma_0/2)\ln(8\epsilon/m)+\arctan[\sigma_0(1-2a^2\epsilon/m)]=-\theta,
\label{bound-men}
\ee
where $z=\sqrt{ma^2/2\epsilon}$.
This resonance is spread out over an energy range of order
$\Gamma_g \sim me^{-\sqrt{2m\pi a^2/\epsilon}}$ and strongly distort
around the Coulomb center. The resonance is sharply defined state with diverging
lifetime $(\Gamma_g)^{-1}\sim e^{\sqrt{2m\pi a^2/\epsilon}}/m$.
Thus, the resonance is practically a bound state.

 The diving point for the energy level defines and depends upon the
parameter $\theta$. This diving of bound levels entails a complete restructuring
of the quantum electrodynamics vacuum in the supercritical Coulomb field \cite{003,grrein}.
As a result, the QED vacuum  acquires the charge, thus leading to the concept of a charged
vacuum in supercritical fields due to the real vacuum polarization \cite{003,grrein}. As was shown
in \cite{grrein} the contribution  to the Green's function from the only pole on the second sheet contains the only term associated with the former lowest bound state:
\bb
G_r({\bf r}, {\bf r'}; E)=i\frac{\Gamma_g\Theta(-m-E)}{(E-E_0)^2+\Gamma_g^2/4}\psi^{cr}_0({\bf r})[\psi^{cr}_0({\bf r'})]^{\dagger},
\label{grsup}
\ee
where $\Theta(z)$ is the step function and $\psi^{cr}_0({\bf r})$ is the ground state of the Dirac Hamiltonian at $a=a_{cr}$ (the critical state) with energy $E_0$ within the gap $-m\leq E_0<m$ but close to $-m$. The  critical  charge $a_{cr}$ is defined as the condition for the appearance of the imaginary part of ``the energy''.  It is important that the Green function of the type (\ref{grsup})
 eliminates the lack of stability of  neutral vacuum for $a>a_{cr}$ (see, \cite{grrein}).
Then, the real vacuum polarization charge density can be determined by
\bb
 j_0^{real}({\bf r})\equiv -\frac{e_0}{2}\int\limits_{R}\frac{dE}{2\pi i}{\rm tr}G({\bf r}, {\bf r'}; E)|_{{\bf r}={\bf r'}}\gamma_0,
\label{cur12}
\ee
where the path $R$ surrounds the  singularity on the unphysical sheet.
Integrating (\ref{cur12}) we obtain $j_0^{real}({\bf r})=-e_0|\psi^{cr}_0({\bf r})|^2$.

We see that the space density of the real vacuum polarization is real quantity
and approximately described with the modulus squared of the fermion wave function
in the critical state:
$$
j_0^{real}(r) \sim - e_0m^2[2(\ln mr)^2-2(\ln mr)/a_{cr} +1/a_{cr}^2], \quad mr\ll 1
$$
and
$$
j_0^{real}(r)\sim -e_0me^{-2\sqrt{r/l}}/r, \quad l=1/\sqrt{2m\epsilon_0}, mr\gg 1,
$$
where $\epsilon_0$ depends upon $a_{cr}$ and the extension parameter $\theta$.

The total induced charge density in massive case with taking into account the real vacuum polarization  (\ref{cur12}) can be estimated as  the sum: $Q_m(r)m^2+j_0^{real}$.

\section{Resume}

In this paper we have studied the vacuum polarization of planar charged  fermions  in
a strong Coulomb potential.
For this we express the density of an induced charge
in the vacuum via the exact Green function, constructed
from solutions of the self-adjoint two-dimensional Dirac Hamiltonians with a strong Coulomb potential.

In the massless case, if the Coulomb center charge is subcritical 
the induced vacuum charge $Q_{ind}$ is localized at the origin  and
has a screening sign, leading to a decrease of the effective Coulomb
center charge. If the Coulomb center charge is supercritical,
the Green function has a discontinuity in the complex plane of ``energy"
due to the singularities on the negative energy axis, which are situated
on the unphysical sheet and related to
the creation of infinitely many quasistationary fermionic states with negative
energies; the induced vacuum charge also has a screening sign but it has a power law form,
causing a modification of Coulomb law at large distances from the Coulomb center.

The finite mass contribution
into the induced charge due to the vacuum polarization is small and insignificantly distorts
the Coulomb potential only at distances of order of the Compton length.
The  induced vacuum charge has a  screening sign.
As is known the quantum electrodynamic vacuum becomes unstable when the Coulomb center charge
is increased from subcritical to supercritical values. In the massive case,
when the Coulomb center charge becomes supercritical
then the lowest state turn into resonance with a diverging lifetime,
which can be described as a quasistationary state with ``complex energy'';
the quantum electrodynamics vacuum acquires the charge due to the so-called real
vacuum polarization. We calculate the real vacuum polarization charge density.
Screening of the Coulomb center charge are briefly  discussed.

We hope  that our results will be helpful for more deep understanding of the fundamental
problem of quantum electrodynamics and can in principle
be tested in graphene with a supercritical Coulomb impurity.

%Screening of the Coulomb center charge due to the vacuum polarization is briefly discussed.
%The problem of the vacuum instability in the supercritical Coulomb field is
%analyzed  by using the self-adjoint Dirac Hamiltonians.
%Since a single electron dynamics in gapless graphene is described by a massless
%two-component Dirac equation  we expect that our results can be used in the
%problem of the vacuum polarization of graphene with a supercritical Coulomb impurity.

%We have calculated and discussed the vacuum polarization induced by the the subcritical and supercritical Coulomb potential in massive case.

\end{document}